\documentclass[prl,superscriptaddress,nofootinbib,twocolumn]{revtex4-1}

\usepackage{graphicx}
\usepackage{color}
\usepackage{epstopdf}
\usepackage{amsmath}

\newcommand{\YH}{$\mathrm{Y}_{\mathrm{H}^+}$} 
\newcommand{\YKr}{\textrm{Y}$_{\textrm{Kr}^+}$} 
\newcommand{\YAr}{\textrm{Y}$_{\textrm{Ar}^+}$} 
\newcommand{\YXe}{\textrm{Y}$_{\textrm{Xe}^+}$} 
\newcommand{\Ytgt}{\textrm{Y}$_{\textrm{NG}^+}$} 
\newcommand{\ON}{\mathrm{ON}}
\newcommand{\OFF}{\mathrm{OFF}}
\newcommand{\BCK}{\mathrm{BCK}}
\newcommand{\Heq}{\mathrm{H}^+} 
\newcommand{\Heqmol}{\mathrm{H}_2^+} 
\newcommand{\TGTeq}{\textrm{NG}^+} 
\newcommand{\Yeq}{\mathrm{Y}}
\newcommand{\intens}{$\times$ 10$^{14}$ W/cm$^2$} 
\newcommand{\Hion}{$\mathrm{H}^+$} 
\newcommand{\Hmol}{$\mathrm{H}_2$} 
\newcommand{\Hmolion}{$\mathrm{H}_2^+$} 
\newcommand{\Kr}{Kr$^+$} 
\newcommand{\waterion}{$\mathrm{H}_2\mathrm{O}^+$} 
\newcommand{\Iest}{$I_{est}$}
\newcommand{\TGT}{\textrm{NG}}
\newcommand{\Kreq}{\textrm{Kr}^+} 

\begin{document}

\title{Precise and accurate measurements of strong-field photoionisation and a transferrable laser intensity calibration standard}

\author{W.C. Wallace}
\author{O. Ghafur}
\author{C. Khurmi}
\author{Satya Sainadh U}
\author{J.E. Calvert}
\author{D.E. Laban}
\affiliation{ARC Centre of Excellence for Coherent X-Ray Science, Griffith University, Brisbane, Queensland, Australia}
\affiliation{Australian Attosecond Science Facility and Centre for Quantum Dynamics, Griffith University, Brisbane, Queensland, Australia}
\author{M.G. Pullen}
\email{Now at: ICFO-Institut de Ci$\grave{e}$ncies Fot$\grave{o}$niques, Av. Carl Friedrich Gauss 3, 08860 Castelldefels (Barcelona), Spain}
\affiliation{ARC Centre of Excellence for Coherent X-Ray Science, Griffith University, Brisbane, Queensland, Australia}
\affiliation{Australian Attosecond Science Facility and Centre for Quantum Dynamics, Griffith University, Brisbane, Queensland, Australia}
\author{K. Bartschat}
\affiliation{ARC Centre of Excellence for Coherent X-Ray Science, Griffith University, Brisbane, Queensland, Australia}
\affiliation{Australian Attosecond Science Facility and Centre for Quantum Dynamics, Griffith University, Brisbane, Queensland, Australia}
\affiliation{Department of Physics and Astronomy, Drake University, Des Moines, Iowa~50311, USA}
\author{A.N. Grum-Grzhimailo}
\affiliation{Skobeltsyn Institute of Nuclear Physics, Lomonosov Moscow State University, Moscow 119991, Russia}
\author{D. Wells}
\author{H.M. Quiney}
\affiliation{ARC Centre of Excellence for Coherent X-Ray Science, University of Melbourne, Melbourne, Victoria, Australia}
\author{X.M. Tong}
\affiliation{Division of Materials Science, Faculty of Pure and Applied Sciences, and Center for Computational Science, University of Tsukuba, 1-1-1 Tennodai, Tsukuba, Ibaraki 305-8577, Japan}
\author{I.V. Litvinyuk}
\affiliation{Australian Attosecond Science Facility and Centre for Quantum Dynamics, Griffith University, Brisbane, Queensland, Australia}
\author{R.T. Sang}
\author{D. Kielpinski}
\email{Corresponding author: d.kielpinski@griffith.edu.au}
\affiliation{ARC Centre of Excellence for Coherent X-Ray Science, Griffith University, Brisbane, Queensland, Australia}
\affiliation{Australian Attosecond Science Facility and Centre for Quantum Dynamics, Griffith University, Brisbane, Queensland, Australia}

\begin{abstract}
Ionization of atoms and molecules in strong laser fields is a fundamental process in many fields of research, especially in the emerging field of attosecond science. So far, demonstrably accurate data have only been acquired for atomic hydrogen (H), a species that is accessible to few investigators. Here we present measurements of the ionization yield for argon, krypton, and xenon with percent-level accuracy, calibrated using H, in a laser regime widely used in attosecond science. We derive a transferrable calibration standard for laser peak intensity, accurate to 1.3\%, that is based on a simple reference curve. In addition, our measurements provide a much-needed benchmark for testing models of ionisation in noble-gas atoms, such as the widely employed single-active electron approximation.
\end{abstract}

\date{\today}

\maketitle
Ionization by strong laser fields drives processes ranging from attosecond pulse generation \cite{Corkum-Krausz-as-rev, Popmintchev-Kapteyn-xray-HHG-rev} to filamentation \cite{Berge-Wolf-filamentation-rev} and remote lasing \cite{Hemmer-Scully-filament-remote-lasing-proposal}. Measurements of strong-field ionization have revealed complex and surprising qualitative features \cite{Grasbon-DeSilvestri-few-cycle-ATI, Wassaf-Maquet-resonant-ionisation-intensity-dependence} that can depend sensitively on the laser intensity.  Precise measurements of strong-field ionization are now being used to probe fundamental physics, such as time delays in photoionization \cite{Eckle-Keller-attoclock-tunnelling-time}, but there is substantial evidence that small systematic offsets in these measurements can obscure the results \cite{Torlina-Smirnova-attoclock-tunneling-time-theory}. In frequency metrology, measurements of atomic transitions are affected by systematic errors arising from the AC-Stark shift and laser intensity uncertainty, thereby limiting the precision of the result \cite{Marian-Ye-2004-time-frequency-spectroscopy-global-structure}. Accurate reference data on strong-field photoionization and laser intensity, especially in the attosecond science regime, are therefore needed for further progress on these questions.

In recent years, our group has used atomic hydrogen (H) to perform quantitatively accurate strong-field measurements that are demonstrably free from systematic errors \cite{Pullen-Kielpinski-few-cycle-H-ionisation, Pullen-Kielpinski-peak-intensity-calibration, Kielpinski-Litvinyuk-strong-field-H-rev}. These measurements are performed in the regime of laser pulse durations and peak intensities that are most widely used in attosecond science. As the simplest electronic system, H has long been recognised as a benchmark species for strong-field physics experiments \cite{Rottke-Welge-hydrogen-low-order-ATI, Paulus-Walther-hydrogen-ATI-expt}. Direct integration of the three-dimensional time-dependent Schr\"odinger equation (3D-TDSE) enables high-accuracy simulation of H and less than 1\% error, with only very minor approximations \cite{GrumGrzhimailo-Bartschat-H-ionization}. Hence, the accuracy of the H data can be certified by their agreement with 3D-TDSE simulations.

Here we use these techniques to perform accurate measurements of the strong-field ionization yield from three commonly used noble-gas targets, and to derive a transferrable, high-accuracy calibration standard for laser intensity. Our data enable accurate inter-comparisons of data taken at various strong-field laboratories; and improved simulations of complex phenomena involving strong-field ionisation, such as filamentation, high-harmonic generation, and laser-induced electron diffraction. We use the noble-gas data to derive a calibration standard for laser peak intensity, given by a simple reference curve, that offers an order of magnitude better accuracy than previous transferrable standards \cite{Smeenk-Staudte-2011-intensity-calibration,Micheau-Lin-2009-intensity-calibration}. Our intensity calibration standard applies to an intensity regime that is readily transferrable to laboratories using few-cycle 800 nm laser systems, including most attosecond science laboratories.

The experimental apparatus is detailed in \cite{Pullen-Kielpinski-few-cycle-H-ionisation, Pullen-Kielpinski-peak-intensity-calibration}. It consists of a well-collimated atomic H beam skimmed from the output of an RF discharge dissociator, which intersects the focus of an intense few-cycle laser. The flux of our custom-constructed atomic source is several orders of magnitude higher than commercial sources \cite{Pullen-Kielpinski-H-ionisation-THESIS}, facilitating a high signal level. The laser generates pulses of 5.5 fs duration (measured at full-width at half-maximum of intensity) with a central wavelength of 800 nm. Ions are created in the overlap region between the atomic beam and the laser beam, and are detected with an ion time-of-flight (ion-TOF) mass spectrometer. A microchannel plate (MCP) located at the end of the ion-TOF detects the ions and outputs a voltage proportional to the ion yield (see Supplementary Information). The overlap region is well-defined, allowing us to account accurately for focal-volume averaging (FVA) effects \cite{Kielpinski-Litvinyuk-strong-field-H-rev}.

\begin{figure}[t]
  \centering
  \includegraphics[width=0.8\columnwidth]{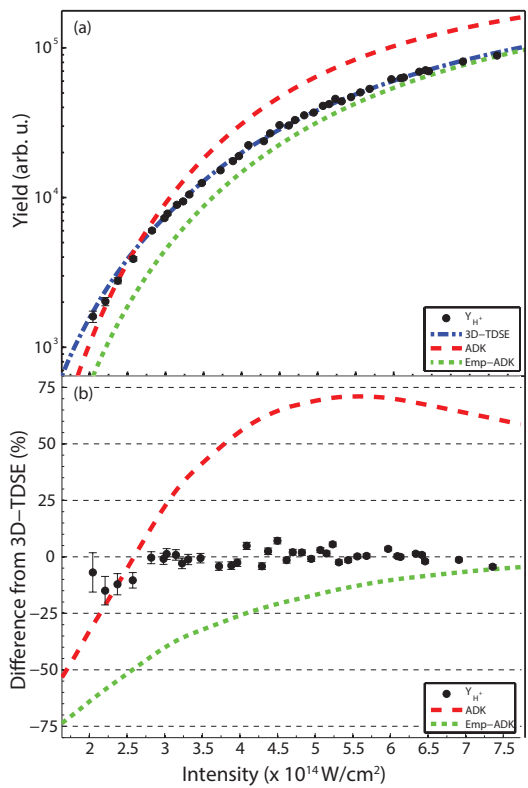}
  \caption{(a) Experimental data for \Hion~yield (solid circles), compared with theoretical predictions from 3D-TDSE (dot-dashed line), standard ADK (dashed line), and empirically corrected ADK (Emp-ADK; dotted line) models. In some cases, the error bars on the data are smaller than the symbols; see Supplementary Information for error estimates. (b)~Percentage difference of experimental \Hion~yield data and both ADK theories from the 3D-TDSE simulations. \vspace{-2 mm}}\label{fig:h-data}
\end{figure}

The measured yield of \Hion~ions resulting from ionisation of atomic H over a range of laser peak intensities is shown in Fig. \ref{fig:h-data}. This yield, denoted \YH, is accurately measured by removing contributions arising from ionisation of undissociated $\mbox{H}_2$ in the beam, and background $\mbox{H}_2\mbox{O}$ vapour. These contaminant signals can be as large as 9.5\% of the desired \YH, and hence must be removed to obtain percent-level accuracy. Errors in \YH~accumulate from MCP voltage baseline subtraction, from short- and long-term laser drifts, and from uncertainties in determining the dissociation fraction. A detailed account of the analysis and error estimation is given in the Supplementary Information.

We certify the accuracy of our data by comparison with accurate 3D-TDSE calculations. As in our previous work \cite{Pullen-Kielpinski-few-cycle-H-ionisation, Pullen-Kielpinski-peak-intensity-calibration}, we perform FVA over the interaction volume, assuming a Gaussian laser profile and a molecular beam of diameter $d$, where $d$ is much smaller than the laser Rayleigh range and much larger than the transverse size of the laser beam. Under these conditions, the FVA becomes independent of beam propagation effects \cite{Kielpinski-Litvinyuk-strong-field-H-rev}. We perform a weighted least-squares fit of the 3D-TDSE yield predictions to the \YH~data using a function of the form
\begin{equation}\label{eq:predH}
P_{\mathrm{H}^+}(I_{est}, A, \eta_1) = A\cdot S(\eta_1 I_{est}).
\end{equation}
\noindent Here $A$ and $\eta_1$ are the fit parameters, while $S$ is the focal-volume averaged and carrier-envelope-phase averaged theory model evaluated at the actual laser intensity $I_0 = \eta_1 I_{est}$. The independently estimated intensity \Iest~is obtained from measurements of the laser parameters -- waist size $w_0$, average power $P$, pulse duration $\tau_p$, and repetition rate $f_{rep}$ -- via
\begin{equation}
\label{eqn:Iest}
I_{est} = \frac{2P}{\pi w_0^2}\frac{1}{f_{rep}\tau_p}.
\end{equation}
The $\eta_1$ fit parameter is a rescaling coefficient of the laser intensity that accounts for the error in \Iest, and permits the accurate retrieval of $I_0$.
The $A$ fit parameter rescales the yield to account for both the unknown atomic density and detector efficiency, but is irrelevant for intensity calibration.

Figure~\ref{fig:h-data} illustrates the agreement between the \YH~data and the 3D-TDSE simulations, certifying that our data are free of systematic error to within our measurement precision of 2\%. A value of $\eta_1$~=~0.641~$\pm$~0.007 is found, indicating that we can calibrate the laser peak intensity to a theory-certified accuracy of better than 1.1\% without systematic error. Normalized residuals from the fit are also shown as percentage deviations from the 3D-TDSE predictions in Fig.~\ref{fig:h-data}. The deviation of $\eta_1$ from its ideal value of 1 is common in high-field experiments, where the uncertainty in \Iest~can easily approach 50\%. We note that if the background is not removed, $\eta_1$ shifts by 3.1\%, much more than our accuracy of 1.1\%.

Our data are easily able to discriminate between the 3D-TDSE and other commonly used theoretical approximations. A well-known alternative to solving the TDSE is the analytic theory of Ammosov, Delone and Krainov (ADK) \cite{Ammosov-Krainov-ADK-ionisation-rate-CLASSIC}. Standard ADK theory is intuitive and straightforward to calculate, but it is known to fail at intensities near to or exceeding the onset of barrier-suppressed ionization (BSI) \cite{Spielmann-Krausz-few-cycle-HHG}. An empirical correction \cite{Tong-Lin-empirically-corrected-ADK} has since been developed to extend the validity of ADK rates to higher intensities. Standard ADK and empirically corrected ADK rates, as well as percentage residuals have also been plotted in Fig.~\ref{fig:h-data} using the $A$ and $\eta_1$ fit parameters obtained from the 5.5 fs 3D-TDSE fit. Standard ADK deviates from the data at high intensities by almost a factor of two, whereas empirically corrected ADK is accurate at the 10\% level there. Nevertheless, the latter model is still clearly ruled out by the data at lower intensities.

\begin{figure*}[t]
  \centering
  \includegraphics[width=2\columnwidth]{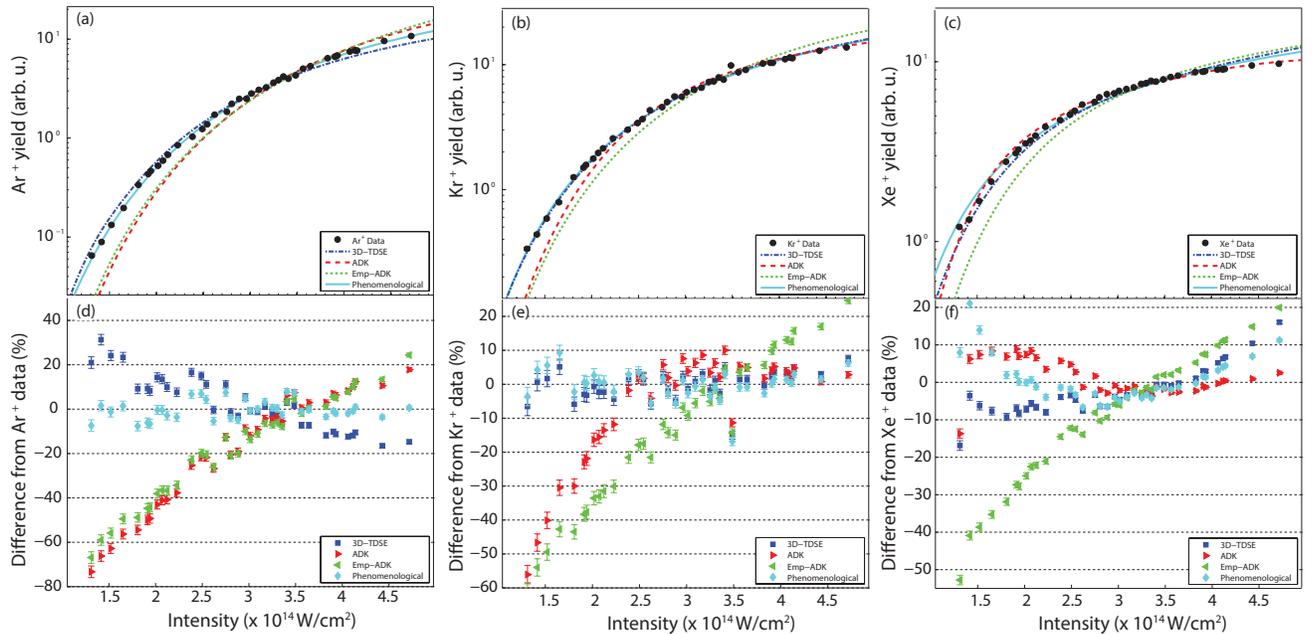}
  \caption{(Color online) Top row: intensity-calibrated experimental data (black circles) for (a) Ar$^+$, (b) Kr$^+$, and (c) Xe$^+$ compared with theoretical predictions from 3D-TDSE (blue, dot-dashed line), ADK (red, dashed line), and Emp-ADK (green, dotted line) models, as well as the phenomenological model given by Eq.~(\ref{eq:H2analytic}) (cyan, solid line). The only fit parameter is the overall rescaling of yield, $A$. In some cases, the error bars on the data are smaller than the symbols. Bottom row: normalized residuals for (d) Ar$^+$, (e) Kr$^+$, and (f) Xe$^+$, for each theoretical prediction, shown as percentage deviations from the experimental yield data for the respective gas targets. \vspace{-5 mm}}\label{fig:h2-data}
\end{figure*}

We now present demonstrably accurate measurements of the photoionization yield of argon, krypton and xenon, providing reference data in a regime for which accurate simulations are not available. The results for the yield of each gas target, denoted \YAr, \YKr, and \YXe are shown in Fig.~\ref{fig:h2-data}. The agreement between theory and experiment for H certifies the accuracy of the noble-gas yields, as these measurements were performed in the same apparatus, using identical laser parameters.

Theoretical ionisation probabilities for the noble-gas atoms are obtained by solving the 3D-TDSE under the single-active electron approximation with the second-order-split operator method in the energy representation \cite{Tong-Toshima-2006-Phase-dependent-ATI,Tong-Chu-1997-HHG-pseudospectral-method}. The model potentials \cite{Tong-Lin-empirically-corrected-ADK} are calculated by using the density functional theory with self-interaction correction \cite{Tong-Chu-1997-DFT}, from which the calculated atomic ionisation potentials are in good agreement with the measured ones. The theoretical simulations are subjected to FVA for comparison with experimental data. As with the atomic H data, we also compare standard ADK and empirically corrected ADK rates with the noble-gas data.

Each of the theory models is compared to \Ytgt (where the subscript NG denotes one of the noble gases) using the calibrated intensity $I_0$ and the fitting method Eq.~(\ref{eq:predH}). The fits and residuals for each target are shown in Fig.~\ref{fig:h2-data}. Since the intensity is already calibrated by the \YH~data, the calibration factor $\eta_1$ is fixed to a value of 1, whilst $A$ is allowed to vary in order to account for the unknown gas~density. While the data are accurate at the 2\% level, the theoretical predictions agree with the data only at the tens of percent level, with both ADK rates performing poorly. These data therefore pose a direct challenge to current models which are widely used to predict results from strong-field ionization experiments.

While all three theories disagree with the experimental data, Fig. \ref{fig:h2-data} shows that we can achieve good agreement with the data using the following phenomenological model
\begin{equation}\label{eq:H2analytic}
  P_{\TGTeq}^{(2\mathrm{D})}(I_{est};A, \eta_2) = A\cdot S_{phenom}(\eta_2 I_{est}),
\end{equation}
\noindent where
\begin{equation}\label{eq:H2phenom}
  S_{phenom}(\eta_2 I_{est}) = \frac{\exp \Big (-\alpha \big(\eta_2 I_{est}/ I_c\big)^{-1/2} \Big )} {1+\big(\eta_2 I_{est}/I_c\big)^{\gamma}}.
\end{equation}
\noindent Here $A$ and $\eta_2$ are the same fit parameters as described in Eq.~(\ref{eq:predH}). The coefficients $\alpha$ and $\gamma$ are set by fitting Eq.~(\ref{eq:H2phenom}) to the 3D-TDSE for each gas target. The value of $I_c$ was determined from the data of Fig. \ref{fig:h2-data} by fixing $\eta_2$ to a value of 1, and substituting \Iest~with our $I_0$ values obtained from the \Hion~fit. The values of these parameters are shown in Table~\ref{tab:fitparams} for each gas target. Our values for $I_c$ include the uncertainty in the \Hion~calibration as well as the fit error, and demonstrate our ability to calibrate the intensity at the 1.3\%, 1.5\%, and 2.5\% levels using Ar, Kr, and Xe as gas targets respectively\footnotemark[1]. The value of $I_c$ is insensitive against the removal of any individual data point from the fit dataset, indicating that the model robustly represents the data over the entire intensity range. However, it is important to note that the phenomenological model is introduced purely as a convenience, so that the reader can easily carry out intensity calibration with a closed-form analytic fit function. We emphasize that Eq.~(\ref{eq:H2phenom}) is not associated with any model of the ionization physics. Hence, we do not expect it to be valid outside the range of intensities studied here.

\begin{table}
\caption{\label{tab:fitparams}Fit parameters used in Eq.~(\ref{eq:H2phenom}) for the Ar$^+$, Kr$^+$, and Xe$^+$ gas targets.}.
\begin{ruledtabular}
\begin{tabular}{cccc}
Fit Parameter& Ar$^+$ & Kr$^+$ & Xe$^+$ \\ \hline
$\alpha$ (arb. units) & 2.84 & 4.24 & 3.71\\
$\gamma$ (arb. units)& -3.03 & -2.49 & -2.69\\
$I_c$ (\intens) & 3.86 $\pm$ 0.05 & 2.06 $\pm$ 0.03 & 1.18 $\pm$ 0.03 \\
\end{tabular}
\end{ruledtabular}
\end{table}

Equation (\ref{eq:H2phenom}) enables absolute intensity calibration at the 1.5\% level for few-cycle 800 nm laser systems, like those widely used for attosecond science. The calibration only requires a mass spectrometer, a few-cycle laser at 800 nm, and a source of either Ar, Kr, or Xe. The intensity range covered by our calibration, \hbox{1~--~5~$\times~10^{14} \:\mbox{W}/\mbox{cm}^2$}, is used by most atomic and molecular strong-field physics experiments, particularly attosecond science experiments. Instructions for using our calibration are detailed in the Supplementary Information.

Our transferrable calibration standard can be shown to reliably determine the \emph{absolute} intensity without systematic errors; in other words, the retrieved intensity can be accurately expressed in the SI unit of $\mbox{W}/\mbox{m}^2$. Previously presented transferrable intensity calibration methods \cite{Micheau-Lin-2009-intensity-calibration, Larochelle-Chin-ADK-ionisation-yield-comparison, Litvinyuk-Corkum-aligned-molecule-ionisation} relied on theoretical approximations whose systematic errors were not fully quantified. The removal of systematic errors, i.e., offsets between the measured value and the ``true value'' of the measured quantity \cite{BIPM-guide-uncertainty-measurement}, is crucial for accurate measurement. Therefore, while these previous methods provide \emph{relative} calibrations of the intensities in the interaction region, their relationship to the SI system, or any other standard system of units, remains unclear.

Our calibration is relatively insensitive to laser parameters other than peak intensity. As shown in our previous work \cite{Pullen-Kielpinski-peak-intensity-calibration}, variations of the pulse duration by 10\% may cause a rescaling of the overall yield, but cause $<1$\% shifts in the retrieved intensity. Similarly, the calibration is not overly sensitive to the precise form of the beam profile: we achieve good theory-experiment agreement for H with beam $M^2$ factors as high as 1.5. In most experiments with molecular beams, including ours, the focal volume averaging is independent of $M^2$ so long as the transverse intensity distribution is Gaussian and is constant within the interaction region. As long as these conditions are satisfied, Eq.~(\ref{eq:H2phenom})~is expected to hold even for much larger values of $M^2$. While the calibration can presently only be used for wavelengths near 800 nm, our simulations show that changes of the wavelength by 50 nm affect the retrieved intensity by $<1$\%. This lack of sensitivity is expected since our laser bandwidth is $>200$ nm. Finally, the calibration can tolerate laser pulse energy fluctuations of at least 0.7\% (root-mean-square), as independently measured on a photo\-diode. We achieve good theory-experiment agreement for H at this level of fluctuation.\\
\indent Other laboratories can transfer our calibration standard to lasers of widely different pulse durations or wavelengths, without the use of an H source, by the following procedure. (i) Calibrate the intensity of a standard few-cycle laser near 800 nm by measuring one of Ar$^+$, Kr$^+$, or Xe$^+$ photoion yields. (ii) Measure the ratios of the beam parameters used in Eq.~(\ref{eqn:Iest}) between the new and standard laser. These ratios can be measured much more accurately than the parameters themselves. From these ratios, derive an absolute calibration of the new laser's intensity. (iii) Measure Ar$^+$, Kr$^+$, or Xe$^+$ photoion yield as a function of the new laser's intensity. The data for the new laser are known to be accurate, since they are referenced to our data by steps (i) and (ii). Finally, (iv) construct a phenomenological fitting function for the new data, to be used in the same way as Eq.~(\ref{eq:H2analytic}) above. A method for intensity calibration for apparatus, in which an atomic beam is not employed, has also been provided in the Supplementary Information.\\
\indent We have presented photoionization yield measurements with an accuracy that improves on previous measurements by an order of magnitude. Our data are obtained in a regime of laser pulse duration and intensity that is widely used for attosecond science, and can be used to benchmark measurement techniques in that field. The measurements are certified at the percent level through the observation of theory-experiment agreement for H. Using the noble-gas data presented here, other laboratories can verify the accuracy of their measurements, calibrate their apparatus, and obtain similarly accurate data for other atomic and molecular species. In the meantime, our data provide accurate reference for simulations of strong-field phenomena involving few-cycle ionization. Finally, we have presented a transferrable calibration of laser intensity that provides an order-of-magnitude accuracy improvement. The standard is readily accessible to laboratories using few-cycle 800 nm lasers and can be further transferred to other laser systems, enabling the correct measurement and interpretation of intensity-sensitive phenomena in strong-field ionization.

\begin{acknowledgments}
This work was supported by the United States Air Force Office of Scientific Research under Grant FA2386-12-1-4025 and by the Australian Research Council (ARC) Centre of Excellence for Coherent X-Ray Science under Grant CE0561787. W.C.W., J.E.C., M.G.P., and D.E.L. were supported by Australian Postgraduate Awards. K.B. acknowledges support from the United States National Science Foundation under Grant No. PHY-1430245 and the XSEDE Allocation No. PHY-090031. X.M.T. was supported by a Grant-in-Aid for Scientific Research (No. C24540421) from the Japan Society for the Promotion of Science. Some calculations were carried out by the supercomputer of the HA-PACS project for advanced interdisciplinary computational sciences by exa-scale computing technology, others on Stampede at the Texas Advanced Computer Center.
D.K. was supported by ARC Future Fellowship FT110100513.
\end{acknowledgments}

\footnotetext[1]{Per the recommendations of the Joint Committee for Guides in Metrology \cite{BIPM-guide-uncertainty-measurement}, we have here stated the $1\sigma$ standard deviation for $I_c$, i.e., the 68\% confidence interval. Therefore, on any particular reproduction of our experiment, the retrieved intensity has a 32\% probability of falling outside the 68\% confidence interval for purely statistical reasons.}

%

\pagebreak
\onecolumngrid
\vspace{\columnsep}
\begin{center}
\noindent \textbf{\large Supplementary information: Precise and accurate measurements of strong-field photoionisation and a transferrable laser intensity calibration standard}
\end{center}
\vspace{\columnsep}
\twocolumngrid

\setcounter{equation}{0}
\setcounter{figure}{0}
\setcounter{table}{0}
\setcounter{page}{1}
\makeatletter
\renewcommand{\theequation}{S\arabic{equation}}
\renewcommand{\thefigure}{S\arabic{figure}}
\renewcommand{\bibnumfmt}[1]{[S#1]}
\renewcommand{\citenumfont}[1]{S#1}
%


\noindent This Supplementary Information details the methods used for acquisition and analysis of the data that are presented in this Letter, alongside detailed instructions on how to calibrate the laser peak intensity according to Eq.~(3) of the manuscript.


\section{Data acquisition}

Data are acquired using a home-built ion time-of-flight (ion-TOF) mass spectrometer of standard design. Ions created in the extraction region via the laser--H-beam interaction are accelerated in a 250 V/cm extraction field before passing through a 1.7 mm slit into a 95 mm long field-free drift region, wherein the ions disperse according to their mass-to-charge ratio. The drift region is terminated with a grounded stainless-steel mesh with an open area ratio of 70\%. An 800 V/cm field is used to post-accelerate the ions into a micro-channel plate (MCP) detector (Hamamatsu F9890-31).

For each laser shot, the voltage waveform V(t) from the output MCP (model: Hamamatsu F9890-31, 450 ps resolution) is recorded via a PCIe digitizer card (model: Agilent U1084A, 250 ps resolution). A LabView-based acquisition program sums up each shot-to-shot waveform for a duration of 10 s, corresponding to 1$\times$10$^4$ laser shots. A 10 s acquisition is obtained for each \textit{setting} of the RF discharge dissociator source (discharge ON, discharge OFF, background BCK, and noble gas NG), and for a number of different estimated laser intensities \Iest. For the case when the dissociator is ON, excited species for atomic H and molecular H$_2$ are generated, however the excited state lifetimes are an order of magnitude shorter than the flight time from the source to the laser-matter interaction region. Any metastable atomic H is rapidly quenched by the electric fields present in the apparatus. Depending on the acquisition, the ion yield Y for three different \textit{species} ($\Heq$, $\Heqmol$, NG$^+$) were measured. The overall yield is calculated as:

 \begin{equation}
 \label{eqn:yield}
    \textrm{Y}_{species}^{setting} = \int_{t_1}^{t_2} V(t) \:dt.
 \end{equation}

It is not a requirement that the detection efficiency of the detection system be equal for all species, only that the detection efficiency for a particular ion species is constant across all laser intensities, which it is. However, this condition can be broken, for example, when the detector is saturated or when space-charge effects occur. We operate our detector below saturation; and both our atomic beam density and background gas density in the interaction region are low enough to avoid space-charge effects.

\begin{figure}[t!]
  \centering
  \includegraphics[width=\columnwidth]{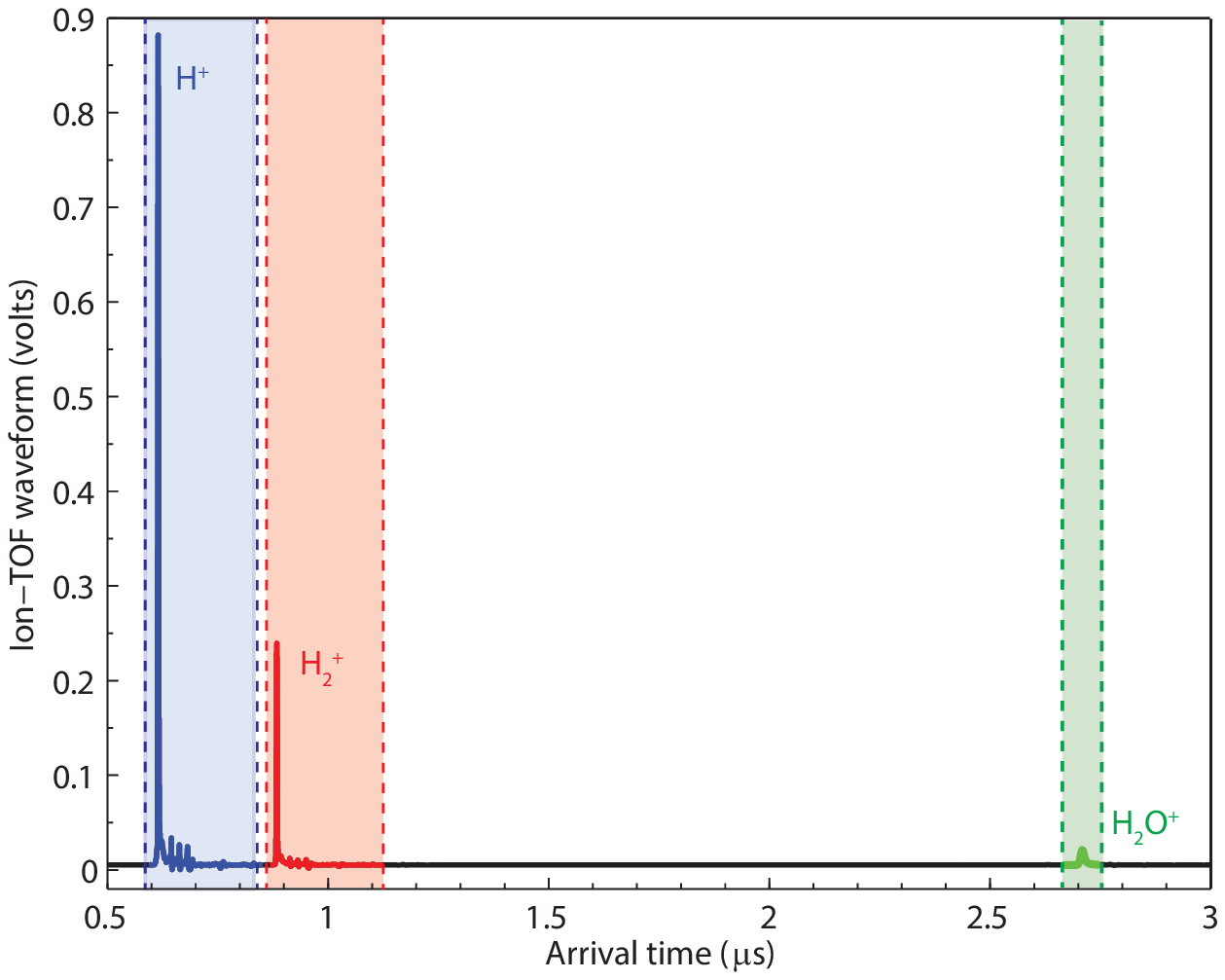}\vspace{-5 mm}
  \caption{Typical ion-TOF waveform obtained with a single laser shot at a calibrated peak intensity of 4.72~$\pm$~0.05~\intens. Shaded regions represent the window of integration for each ion species (\Hion~in blue, \Hmolion~in red, and \waterion~in green). \vspace{-3 mm}}\label{fig:waveform}
\end{figure}

A typical ion-TOF waveform is shown in Fig.~\ref{fig:waveform}. This waveform was obtained with a calibrated peak laser intensity of 4.72~$\pm$~0.05~\intens~and illustrates the presence of three ion species: \Hion, \Hmolion~from residual \Hmol~present in the beam, and \waterion~from background gas in the vacuum chamber. Evaluation of the yield of each ion species is performed by integrating the ion-TOF waveform in the time interval bounded by the shaded regions of interest. Ringing in the waveform is observed, alongside a non-zero baseline. Errors in the evaluation of the yield due to baseline subtraction and ringing are discussed below in Sec. \ref{sec-yield-est}.

To isolate the yield of \Hion~ions resulting from ionisation of atomic H we combine individual waveform measurements as:
\begin{equation}\label{eq:YH}
  \Yeq_{\Heq} = \Yeq_{\Heq}^{\ON} - \Yeq_{\Heq}^{\BCK} - (1 - \mu)(\Yeq_{\Heq}^{\OFF} - \Yeq_{\Heq}^{\BCK}).
\end{equation}
Here the superscript denotes the setting of the discharge (BCK for when the \Hmol~gas is off, OFF for when the \Hmol~gas is on but the discharge is off, and ON for when the discharge is on); and the subscript denotes the ion species. The $\mu$ term represents the dissociation fraction and is the percentage of \Hmol~molecules dissociated when the H discharge source is turned ON. It is given by:
\begin{equation}\label{eq:mu}
  \mu = 1 - \frac{\Yeq_{\Heqmol}^{\ON} - \Yeq_{\Heqmol}^{\BCK}}{\Yeq_{\Heqmol}^{\OFF} - \Yeq_{\Heqmol}^{\BCK}}.
\end{equation}

Determining the yield from the noble gas is much simpler. For example, the \Kr~ion yield (denoted \YKr) from photoionisation of krypton  atoms was determined by combining the NG and BCK measurements according to:
\begin{equation}\label{eq:YH2}
  \Yeq_{\Kreq} = \Yeq_{\Kreq}^{\TGT} - \Yeq_{\Kreq}^{\BCK},
\end{equation}
\noindent where Kr denotes that krypton gas (rather than \Hmol) flows through the RF dissociator source. The discharge setting is OFF when measuring noble-gas yields.

\section{Error analysis}

The overall uncertainty in the photoionisation yield measurements has four error contributions:
\begin{enumerate}
\item Estimation of yield from the MCP voltage waveform
\item Short-term drift (within an individual measurement)
\item Long-term drift (between the ON, OFF, and BCK measurements)
\item Error in the dissociation fraction estimate entering Eq.~(\ref{eq:YH}).
\end{enumerate}
These contributions are summarised in Fig.~2 of the main text. The error in yield estimation arose from background subtraction and ringing. Short-term drift errors were calculated via a modified Allan deviation analysis \cite{Howe-Barnes-modified-allan-deviation}. Long-term drift errors were estimated from measurements of the power spectral density (PSD) of the noise \cite{Harris-Ledwidge-noise-analysis-BOOK}. Errors in the dissociation fraction were inferred from the errors in the yield measurements. The long-term drift was the dominant error source, with a factor of four higher contribution than the other error sources.

\subsection{Yield estimation}
\label{sec-yield-est}

The raw waveform acquired from the multichannel plate has a non-zero baseline, owing to a small but non-negligible DC voltage offset at the input to the PCIe digitizer card. The voltage baseline was removed by averaging the waveform over a 550 ns time interval prior to the onset of the first ion peak, and subtracting this average from the entire waveform. The error in the baseline subtraction was calculated using the standard deviation of the corrected baseline, and is denoted $\sigma_B(I)$.

\begin{figure}[t!]
        \centering
        \includegraphics[width=0.98\columnwidth]{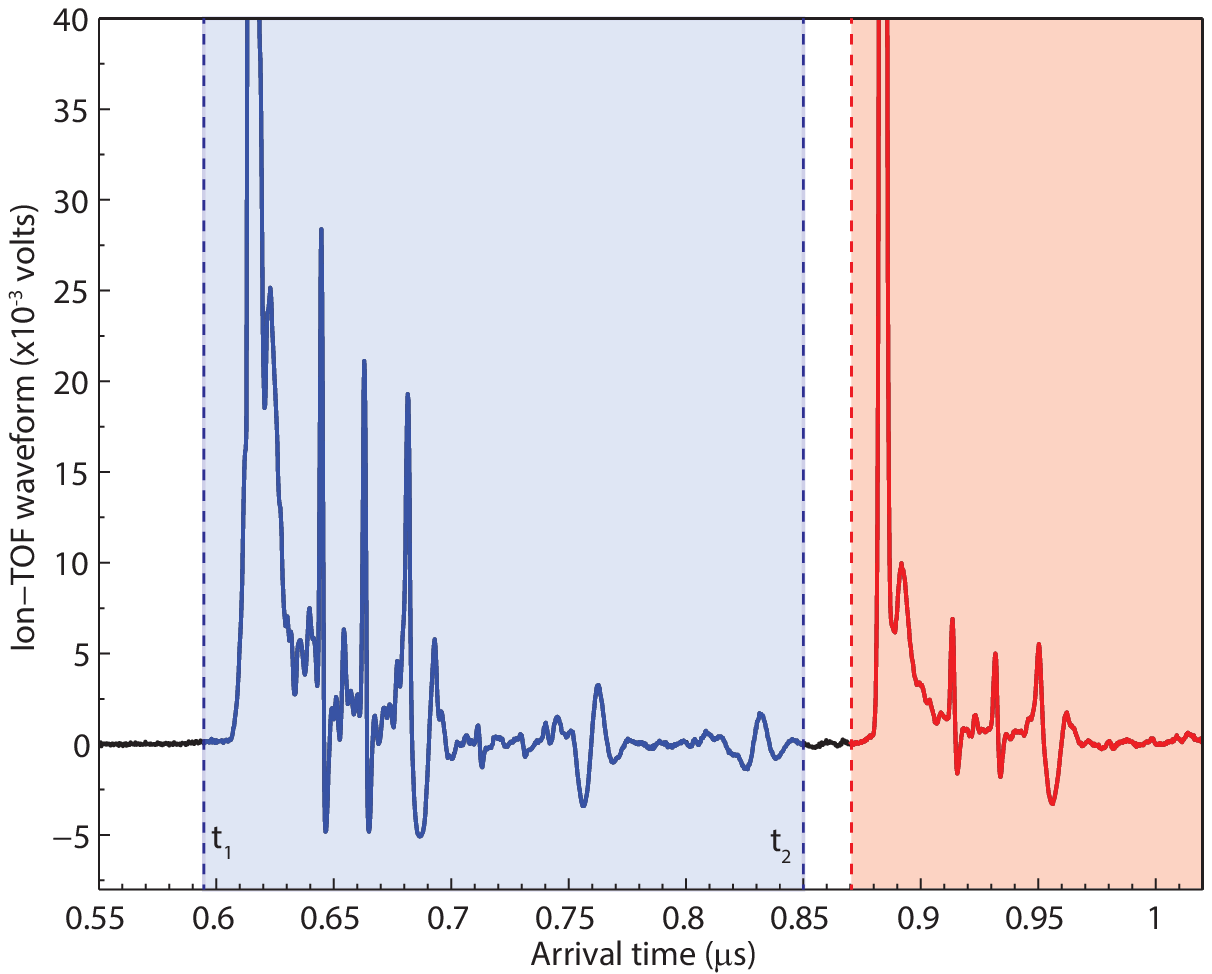}
        \caption{\label{fig:ringing} Example of waveform ringing at high laser peak intensities. The ringing extends from the end of the H$^+$ peak (blue) until the onset of the \Hmolion peak (red). The time interval bounded by the blue box is on the order of 250 ns. Increasing the width of the time interval does not affect the overall H$^+$ yield at the 0.05\% level from one half-cycle of the ringing to the next.}
 \end{figure}

Waveform ringing was observed at high laser peak intensities, as illustrated in Fig. \ref{fig:ringing}. This ringing is due to a small impedance mismatch between the MCP output and the PCIe digitizer input. The ringing constitutes a real and observable output current, and therefore contributes to the overall yield signal for each ion. The time interval over which each ion peak was integrated is chosen such that there is less than 0.05\% difference in the overall yield when the time interval is extended by one-half cycle of the ringing. The residual error is negligible in comparison to the other contributions. Hence, we discard its contribution to the overall error in the yield, but for completeness still present it here.

\subsection{Short-term drift}
\begin{figure}[t!]
        \centering
        \includegraphics[width=\columnwidth]{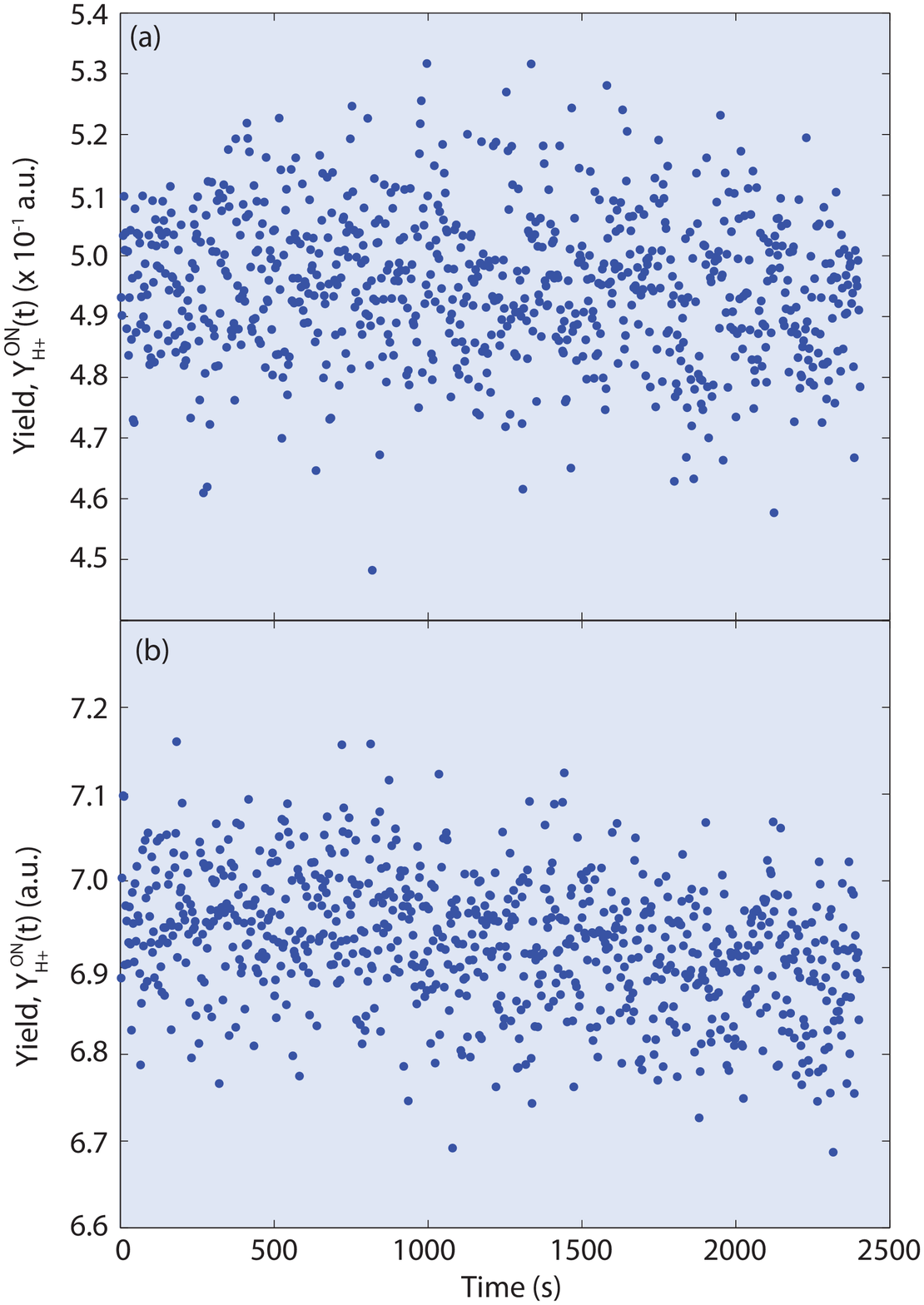}
        \caption{\label{fig:timeseries} Time series measurements of ion yields for  $\Yeq_{\Heq}^{\ON}$ acquired at laser peak intensities of (a)~1.7 and (b) 3.9~\intens. Yield units are arbitrary, but are kept consistent between the two datasets.}
 \end{figure}

We use the IEEE-recommended \cite{IEEE-freq-time-metrology-standard} method of overlapping Allan variance analysis \cite{Howe-Barnes-modified-allan-deviation} to measure the short-term variance in the time domain. To quantify the variance we first recorded 800 consecutive ion-TOF waveforms, each averaged over 3000 laser shots, for a total time interval of approximately 2400 seconds. During this time interval, no experimental parameters were adjusted. Next, the yield of each ion species was extracted from each ion-TOF waveform via the baseline subtraction and integration method outlined earlier. The resulting time series measurements of ion yield versus time, Y$_{species}^{setting}$(t), taken at low ($I_L$~=~1.7~\intens) and high ($I_H$~=~3.9~\intens) peak laser intensities, are shown in Fig.~\ref{fig:timeseries} for the case of $\Yeq_{\Heq}^{\ON}$. The analysis described below was carried out also for all the noble-gas yields, but here we only discuss $\Yeq_{\Heq}^{\ON}$ measurements for simplicity.

\begin{figure}[t!]
        \centering
        \includegraphics[width=\columnwidth]{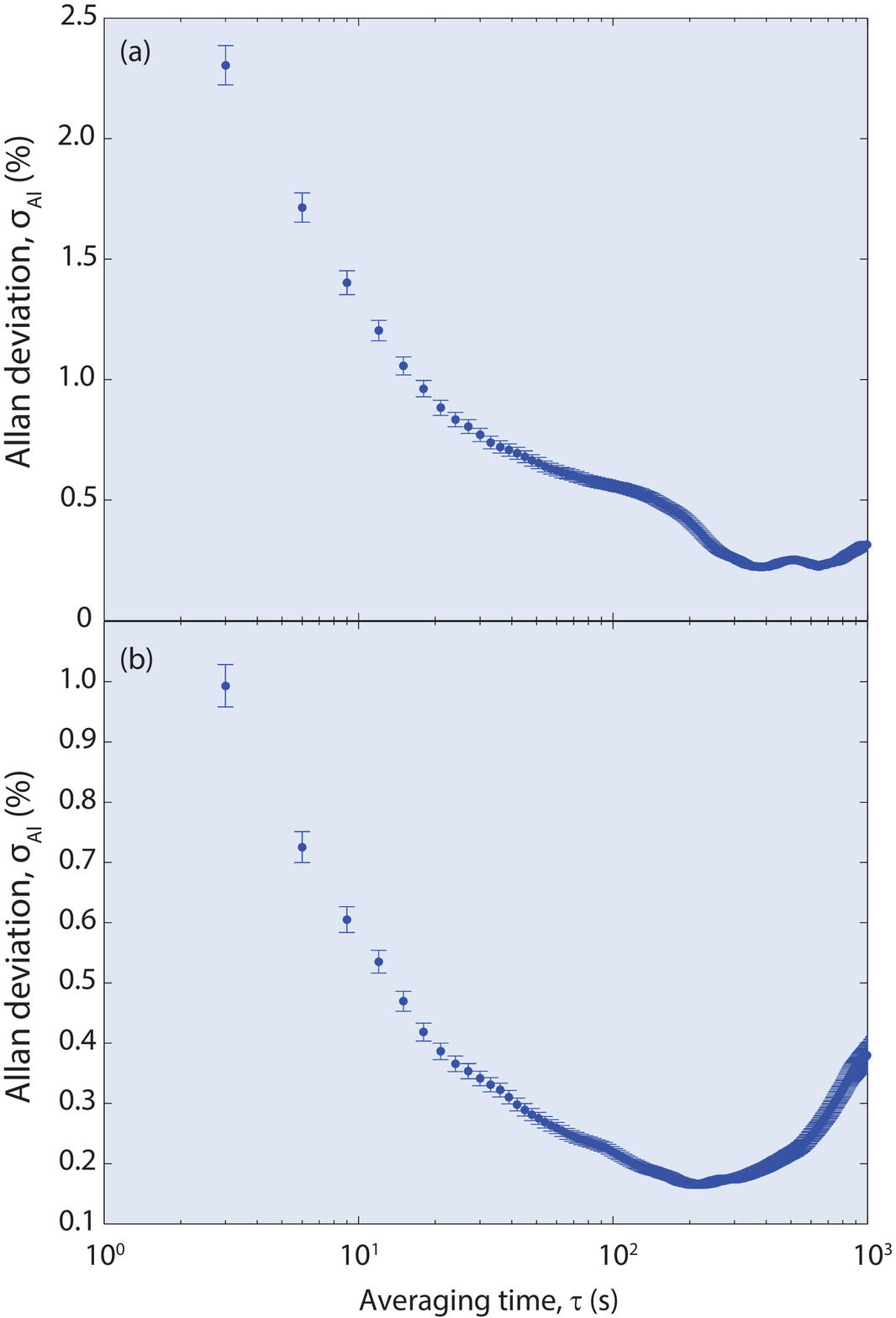}
        \caption{\label{fig:allan} Overlapping Allan deviation shown as percentage error in $\Yeq_{\Heq}^{\ON}$ at laser peak intensities of (a)~1.7 and (b) 3.9~\intens.}
 \end{figure}

 We perform Allan variance analysis on both low- and high-intensity time series measurements according to the method given in \cite{Howe-Barnes-modified-allan-deviation}. The result of this analysis is shown in Fig.~\ref{fig:allan}. The \textit{y}-axis shows the percentage error incurred for an averaging time of $\tau$ seconds. In our experiment, the $\tau$ = 10 s averaging time for a single ion-TOF waveform acquisition corresponds to a measured Allan deviation error, $\sigma_{Al}$, in $\Yeq_{\Heq}^{\ON}$ of less than 1.4\% (0.6\%) for low (high) laser peak intensity. Ideally, one would like to measure the intensity-dependent Allan deviation, $\sigma_{Al}(I)$, across all intensities, but as seen below, the short-term error contribution is negligible in comparison to the long-term error contribution.
\subsection{Long-term drift}
\begin{figure}[t!]
        \centering
        \includegraphics[width=\columnwidth]{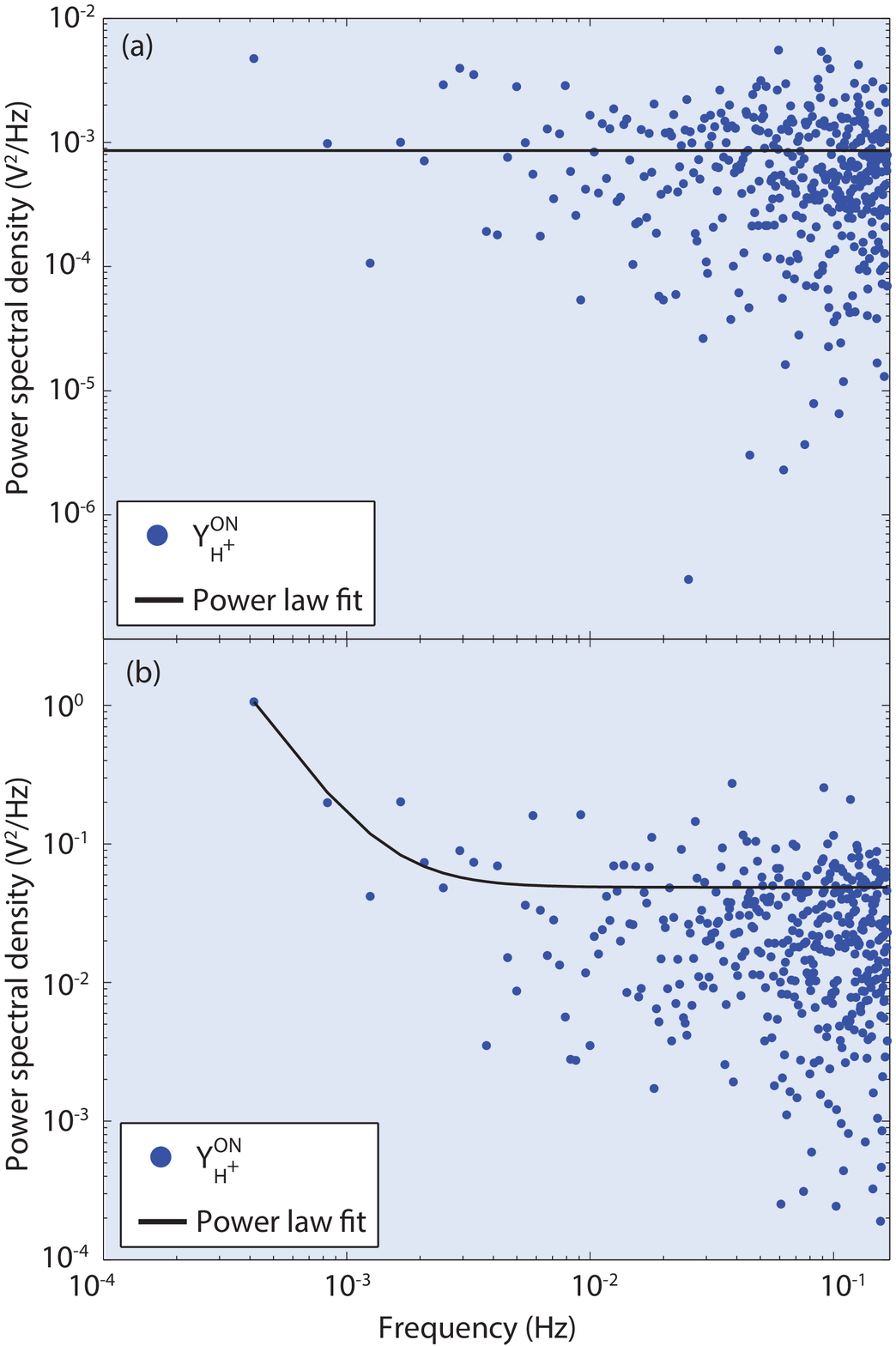}
        \caption{\label{fig:psd} Power spectral densities calculated from the time series measurement of $\Yeq_{\Heq}^{\ON}$ acquired at laser peak intensities of (a)~1.7 and (b) 3.9~\intens. The power law fits from Eq.~(S\ref{eqn:powerlaw}) are shown as solid black lines.}
 \end{figure}

Errors due to long-term drifts are estimated via power spectral density (PSD) analysis \cite{Harris-Ledwidge-noise-analysis-BOOK} of the $\Yeq_{\Heq}^{\ON}$(t) time series measurements. Briefly, the PSD of a time series or signal gives the distribution of power contained within that signal per unit frequency. The PSD $S_{\textrm{YY}}(\omega)$ of a time varying signal y(t) is given by:
\begin{equation}
\label{eqn:psd}
S_{\textrm{YY}}(\omega) = \lim_{T \to \infty} \frac{1}{2T}|\textrm{Y}_T(\omega)|^2,
\end{equation}
\noindent where $\textrm{Y}_T(\omega)$ is the Fourier transform of y(t):
\begin{equation}
\label{eqn:fft}
\textrm{Y}_T(\omega) = \int_{-T}^T \textrm{y(t)}\mathrm{e}^{-i \omega t} \:dt.
\end{equation}
\noindent In practice, Eq.~(S\ref{eqn:fft}) is true only if y(t) is a signal with zero mean, such that:
\begin{equation*}
\label{eqn:timeseriesmean}
\textrm{y(t) = Y(t)} - \langle \textrm{Y(t)} \rangle.
\end{equation*}
The calculated PSD for the $\Yeq_{\Heq}^{\ON}$(t) time series measurements is shown in Fig.~\ref{fig:psd} on a log-log plot. As expected, the low-frequency PSD at laser peak intensity $I_H$ follows a power law, which takes the form of:
\begin{equation}
\label{eqn:powerlaw}
S_{\textrm{YY}}(f,I_H) = \frac{A(I_H)}{f ^\gamma} + B(I_H),
\end{equation}
\noindent where $A(I_H) = 5.97 \pm 0.38 \times 10^{-9}$ V$^2$/Hz$^{\gamma-1}$, $B(I_H) = 4.86 \pm 0.16 \times 10^{-3}$ V$^2$/Hz and $\gamma = 2.43 \pm 0.08$ (a.u.). At low laser peak intensity $I_L$, we find that $S_{\textrm{YY}}(f,I_L)$ is purely white noise with $B(I_L) = 8.42 \pm 0.43 \times 10^{-4}$ V$^2$/Hz and $A(I_L)$ consistent with zero. Using the power-law fit to the PSD enables us to interpolate between the discrete sample frequencies provided by the time-series data. Again, a similar analysis was performed on the noble-gas yield data but for the sake of brevity is not presented here.
A reliable estimate of the long-term drift error $\sigma_P$ is found by integrating the PSD over the frequency range relevant to the time required to perform our measurements \cite{Harris-Ledwidge-noise-analysis-BOOK}:
\begin{equation}
\label{eqn:psdvariancelimits}
\sigma_{P}^2 = 2\pi \int_{f_1}^{f_2} S_{\textrm{YY}}(f) \mathrm{d}f.
\end{equation}
\noindent Here $f_1$ is the frequency corresponding to period required to acquire all ion-TOF waveforms, and $f_2$ is the period required for a single ion-TOF waveform. In our measurements, $f_2 \leq f_c$, where $f_c$ is the frequency below which the $B(I)$ contribution to the PSD is negligible. Thus the $B(I)$ term in Eq.~(S\ref{eqn:powerlaw}) has minimal effect on the measured variance. We measure $\sigma_P$ in $\Yeq_{\Heq}^{\ON}$ to be 1.4\% (0.6\%) for low (high) laser peak intensity.

Ideally, one would like to measure the long-term drift errors at each laser intensity independently, but such measurements are impractical owing to their time-consuming nature. Consideration of the different experimental parameters reveals that the absolute contributions (as opposed to relative contributions) to the error scale most quickly with fluctuations in atomic beam density and least quickly with fluctuations in laser peak intensity. From these observations, and the PSD calculated at $I_L$ and $I_H$ previously, we develop a general error model which provides the best estimate of the upper-bounds on $A(I)$ and $B(I)$ for use in calculating the intensity-dependent PSD $S_{\textrm{YY}}(f,I)$.

At $I \leq I_L$, $A(I_L)$ is known to be negligible, and in the region of intensities given by $I_L \leq I \leq I_H$, $A(I)$ can scale no faster than $|$dY/dI$|^2$, with these drifts arising from fluctuations in laser intensity. At intensities $I_H \leq I$, $A(I)$ may scale no faster than $|$Y(I)$|^2$, with this scaling arising from drift in inlet gas pressure and dissociation efficiency. We derive similar bounds for $B(I)$, except $B(I \leq I_L)$ is set equal to $B(I_L)$.

From these observations, we obtain the following piecewise upper limits on both $A(I)$ and $B(I)$:

\begin{equation}
\label{eqn:Apiece}
A(I) = \left\{
\begin{tabular}{ l c l }
  0 & $\mspace{20mu}  \mspace{20mu}$ & $I \leq I_L$ \\
  $A(I_H)\Bigg[\frac{\textrm{Y}'(I)}{\textrm{Y}'(I_H)}\Bigg]^2$ & $\mspace{20mu}  \mspace{20mu}$ & $I_L \leq I \leq I_H$  \\
  $A(I_H)\Bigg[\frac{\textrm{Y}(I)}{\textrm{Y}(I_H)}\Bigg]^2$ & $\mspace{20mu}  \mspace{20mu}$ & $I_H \leq I$ \\
\end{tabular} \right.
\end{equation}

\begin{equation}
\label{eqn:Bpiece}
B(I) = \left\{
\begin{tabular}{ l c l }
  $B(I_L)$ & $\mspace{20mu}  \mspace{20mu}$ & $I \leq I_L$ \\
  $B(I_H)\Bigg[\frac{\textrm{Y}'(I)}{\textrm{Y}'(I_H)}\Bigg]^2$ & $\mspace{20mu}  \mspace{20mu}$ & $I_L \leq I \leq I_H$  \\
  $B(I_H)\Bigg[\frac{\textrm{Y}(I)}{\textrm{Y}(I_H)}\Bigg]^2$ & $\mspace{20mu}  \mspace{20mu}$ & $I_H \leq I$, \\
\end{tabular} \right.
\end{equation}

\noindent where Y$^{\prime}(I)$ = dY/d$I$.

We note that for $B(I)$, shot-noise can be considered as a scaling option for intensities $I_L \leq I \leq I_H$, since the $B(I)$ term of the PSD has the frequency-independent power spectrum characteristic of white noise. This option is not present for the $A(I)$ term since that term is not frequency-independent. However, it turns out that $|$dY/dI$|^2$ scales in nearly the same way as shot noise over the relevant intensity range. Since the contribution of $B(I)$ to the total variance is small in any case, we find that the difference in scaling relations leads to a negligible effect on the long-term drift error.

We calculate the intensity-dependent long-term drift error $\sigma_{P}(I)$ by calculating the intensity-dependent PSD using the piecewise bounding conditions for $A(I)$ and $B(I)$ from Eqs.~(S\ref{eqn:Apiece}) and (S\ref{eqn:Bpiece}) respectively, and substituting them into Eq.~(S\ref{eqn:powerlaw}). The intensity-dependent PSD is subsequently integrated according to Eq.~(S\ref{eqn:psdvariancelimits}) to arrive at $\sigma_{P}(I)$.

\begin{figure}[t]
  \centering
  \includegraphics[width=\columnwidth]{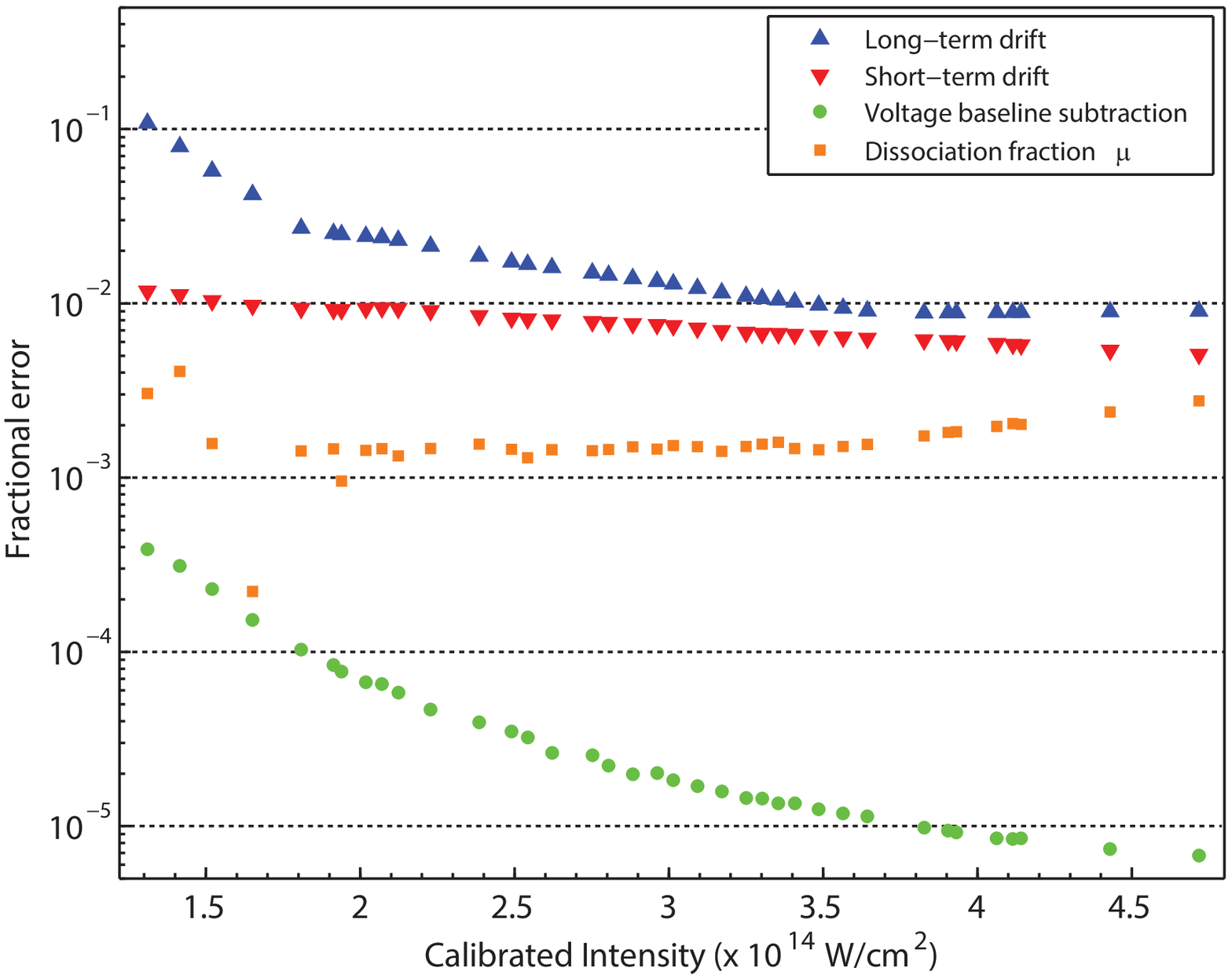}
  \caption{Fractional error contributions for \YH~shown in log scale. The primary contribution is long-term drift error that accumulates over our 1 hour total acquisition time. The kink in long-term drift error arises from our highly conservative modeling of drift error between 1.7 and 3.9 \intens.\vspace{-3 mm}}\label{fig:h-errors}
\end{figure}

All three intensity-dependent errors, $\sigma_{B}(I)$, $\sigma_{Al}(I)$, and $\sigma_{P}(I)$ are combined in quadrature to obtain the overall error in the yield, $\sigma_{Y_{species}^{setting}}$, for a given apparatus setting and ion species (e.g. $\sigma_{Y_{H^+}^{ON}}, \sigma_{Y_{H^+}^{OFF}}$). The final errors in \YH and \YKr, from Eq.~(1) and (3) of the manuscript, respectively, are then propagated through from the errors of each individual $\sigma_{Y_{species}^{setting}}$ term. Figure \ref{fig:h-errors} illustrates the contributions of the errors in the determination of $\Yeq_{\Heq}$.

\makeatletter 
\def\tagform@#1{\maketag@@@{(\ignorespaces#1\unskip\@@italiccorr)}}
\makeatother

\section{Beam propagation considerations}
As stated in the manuscript, we achieve good agreement with $M^2$ values of 1.5, close to the theoretical limit of 1. Two separate measurements of the $M^2$ value were obtained before and after the experimental chamber using a CCD beam profiler, and Gaussian propagation was confirmed. Further beam profile measurements were made with a series of 50 nm bandpass filters extending across the range of our pulse bandwidth (e.g. 700-750 nm, 750-800 nm). Results showed that our beam had Gaussian propagation properties across the entire spectral bandwidth, with negligible effect on the position of the focus in comparison to the Rayleigh range of the beam. As an off-axis parabolic mirror was used to focus the beam (as opposed to a lens), the only diffracting effects that could arise are from the 0.5 mm viewport on our experimental chamber. Any possible effects on the position of the foci for different wavelengths are negligible owing to the thin-ness of the viewport with respect to the 750 mm focal length of the parabolic mirror.

\section{Method for intensity calibration}
This subsection outlines the method required to calibrate the peak laser intensity for any experimental apparatus utilizing a few-cycle laser and krypton gas target.
\subsection{Take the data}
Measure the single ionisation yield from the Kr $\rightarrow \textrm{Kr}^+$ process as a function of $I_{est}$. Calculate \Iest using Eq.~(2) from the manuscript, which we restate for convenience:
\begin{equation}
\label{eqn:Iest}\tag{2}
I_{est} = \frac{2P}{\pi w_0^2}\frac{1}{f_{rep}\tau_p},
\end{equation}
\noindent where $P$ is the laser average power, $w_0$ is the beam waist, $f_{rep}$ is the repetition rate and $\tau_p$ is the pulse duration. As long as the relative intensities are well known, the actual values for \Iest only need to be known roughly, so that the subsequent least-squares fit converges properly. The example $\Yeq_{\Kreq}$ data shown in Table~\ref{tab:Krdata} are actual experimental data obtained using our apparatus.

\subsection{Determine which fit equation is most suitable}
If the Kr target lies well within the Rayleigh range of the laser focus, the appropriate fit equation is Eq. (3) from the manuscript, which we restate for convenience:
\begin{equation}\label{eq:H2analytic}\tag{3}
  \textrm{P}_{\Kreq}^{(2\textrm{D})}(I_{est};A, \eta_2) = A\cdot S_{phenom}(\eta_2 I_{est}),
\end{equation}
\noindent where
\begin{equation}\label{eq:H2phenom}\tag{4}
  S_{phenom}(\eta_2 I_{est}) = \frac{\exp \Big (-\alpha \big(\eta_2 I_{est}/ I_c\big)^{-1/2} \Big )} {1+\big(\eta_2 I_{est}/I_c\big)^{\gamma}}.
\end{equation}
Here $A$ and $\eta_2$ are fit parameters, $\alpha$~=~4.24, \hbox{$\gamma=-2.49$} and $I_c$ =~2.06~$\pm$~0.03~\intens. In this situation, the cylindrically symmetric geometry of the interaction region negates the need for focal-volume averaging (FVA) along the propagation direction of the laser.

\makeatletter 
\def\tagform@#1{\maketag@@@{(S\ignorespaces#1\unskip\@@italiccorr)}}
\makeatother

However, if the gas density is uniform within the entirety of the experimental apparatus, as opposed to an atomic beam in our case, then Eq.~(\ref{eq:H2analytic}) must be modified to allow for integration along the entire length of the laser beam. The predicted yield for this case is:
\begin{equation}\label{eq:H2analytic3D}
  \mbox{P}_{\Kreq}^{(3\mathrm{D})} = \int_0^1 \frac{\mbox{P}_{\Kreq}^{(2\mathrm{D})}(vI_{est})}{v^{5/2}\sqrt{1-v}} \:dv,
\end{equation}
where $v$ = ${w_0}^2/w(z)^2$, and $w(z) = w_0\sqrt{1 + (z/z_\textrm{R})^2}$ is the laser spot size at a distance $z$ along the propagation direction for a laser with beam waist size $w_0$ and Rayleigh range $z_\textrm{R}$. Again, the actual peak intensity of the laser, $I_0$, is given by $I_0 = \eta_2 I_{est}$, and the error in $I_0$ is given by the combined error from $\eta_2$ and $I_c$.

\subsection{Fit the data}
Perform a non-linear weighted least-squares fit of the measured yield data (column 2 of Table \ref{tab:Krdata}) with the appropriate fitting equation using $A$ and $\eta_2$ as fit coefficients. In the fit, the weight of each data point should be set equal to the inverse of the error for that point. Fitting the data in Table \ref{tab:Krdata} to Eq. (3) of the manuscript gives the value $\eta_2$ = 0.638 $\pm 0.008$. As expected, this is nearly identical to the $\eta_1$ obtained from the H$^+$ intensity calibration, as both the H$^+$ and \Kr~ yield data were taken with the same apparatus at approximately the same time.
\begin{center}
    \begin{table}[ht!]
    \begin{tabular}{ | c | c | c | c |}
    \hline
    \Iest & \YKr   & $\sigma_{\textrm{Y}_{\textrm{Kr}^+}}$ & $I_0$  \\
    (x10$^{14}$~W/cm$^2$) &(arb.~units) & (abs. error) & ($\times 10^{14}$~W/cm$^2$) \\ \hline
        2.04    &   0.2349  &0.0064     &1.31\\\hline
        2.21    &   0.3059  &0.0078     &1.42\\\hline
        2.37    &   0.4098  &0.0099     &1.52\\\hline
        2.58    &   0.5542  &0.0125     &1.65\\\hline
        2.82    &   0.8780  &0.0187     &1.81\\\hline
        2.98    &   1.0478  & 0.0215    &1.91\\\hline
        3.03    &   1.1034  & 0.0224    &1.94\\\hline
        3.15    &   1.2377  & 0.0245    &2.02\\\hline
        3.23    &   1.3752  & 0.0268    &2.07\\\hline
        3.31    &   1.4978  & 0.0286    &2.12\\\hline
        3.48    &   1.7985  & 0.0332    &2.23\\\hline
        3.72    &   2.1119  & 0.0370    &2.38\\\hline
        3.88    &   2.3835  & 0.0402    &2.49\\\hline
        3.97    &   2.5669  & 0.0425    &2.54\\\hline
        4.09    &   3.0203  & 0.0486    &2.62\\\hline
        4.29    &   3.1995  & 0.0491    &2.75\\\hline
        4.38    &   3.5130  &0.0529     &2.8\\\hline
        4.50    &   3.8918  &0.0569     &2.88\\\hline
        4.62    &   3.8866  &0.0553     &2.96\\\hline
        4.70    &   4.2184  &0.0591     &3.01\\\hline
        4.82    &   4.4026  &0.0602     &3.09\\\hline
        4.95    &   4.5834  &0.0612     &3.17\\\hline
        5.07    &   5.0966  &0.0663     &3.25\\\hline
        5.15    &   5.1435  &0.0658     &3.3\\\hline
        5.23    &   5.5428  &0.0696     &3.35\\\hline
        5.32    &   5.3139  &0.0655     &3.41\\\hline
        5.44    &   6.9126  &0.0830     &3.49\\\hline
        5.56    &   6.0803  &0.0712     &3.56\\\hline
        5.68    &   6.3627  &0.0726     &3.64\\\hline
        5.97    &   7.1661  &0.0768     &3.83\\\hline
        6.09    &   7.2500  &0.0757     &3.9\\\hline
        6.13    &   7.2485  &0.0750     &3.93\\\hline
        6.34    &   7.7381  &0.0774     &4.06\\\hline
        6.42    &   8.0049  &0.0791     &4.11\\\hline
        6.46    &   7.9000  &0.0777     &4.14\\\hline
        6.91    &   9.0530  &0.0861     &4.43\\\hline
        7.36    &   9.6458  &0.0912     &4.72\\\hline

    \end{tabular}
    \caption {Actual experimental $\Yeq_{\Kreq}$ data as a function of \Iest taken with our apparatus. This data can be used to check the implementation of our intensity calibration procedure.} \label{tab:Krdata}
    \end{table}
\end{center}
\subsection{Compute actual laser intensity}
The true set of laser intensities $I_0$ is given by $\eta_2I_{est}$. The uncertainty in the true laser intensity, $\sigma_{I_0}$, is then given by $\sigma_{I_0}^2 = \sigma_{\eta_2}^2 + \sigma_{I_c}^2$, where the value $\sigma_{I_C}$ = 1.5\% was computed in the main text. From our sample data, $\sigma_{I_0}$ equates to 2.0\%.

\end{document}